# ANDES, the high-resolution spectrograph for the ELT: RIZ Spectrograph preliminary design

Bruno Chazelas[a], Yevgeniy Ivanisenko[a], Audrey Lanotte[a], Pablo Santos Diaz[a], Ludovic Genolet[a], Michael Sordet[a], Ian Hughes[a], Christophe Lovis[a], Tobias M. Schmidt[a], Manuel Amate[f], José Peñate Castro[f], Afrodisio Vega Moreno[f], Fabio Tenegi[f], Roberto Simoes[f], Jonay I. González Hernández[f], María Rosa Zapatero Osorio[h], Javier Piqueras[g], Tomás Belenguer Dávila[g], Rocío Calvo Ortega[g], Roberto Varas González[g], Luis Miguel González Fernández[g], Pedro J. Amado[g], Jonathan Kern[c], Frank Dionies[c], Svend-Marian Bauer[c], Hakan Önel[c], Arto Järvinen[c], Joar Brynnel[c], Christine Füßlein[c], Olga Bellido[c], Jörg Weingrill[c], Domenico Giannone[c], Wolfgang Gaessler[d], Michael lehmitz[d], Adrian Kaminski[e], Ingo Stilz[e], Michael Sigwarth[i], Alessandro Marconi[j,k], Paolo Di Marcantonio[b], Ernesto Oliva[k], Igor Coretti[b], Matteo Aliverti[l], Giorgio Pariani[l], Lorenzo Cabona[l], Edouardo Maria Alberto Radaelli[l], Marcello Scalera[l], Andrea Balestera[l]

[a]Département d'Astronomie, Université de Genève, Chemin de Pegasi 51, Versoix (Switzerland),
[b]INAF - Osservatorio Astronomico di Trieste, via G. B. Tiepolo 11, 34143 Trieste, (Italy),
[c]Leibniz Institute for Astrophysics Potsdam (AIP), An der Sternwarte 16, D-14482 Potsdam(Germany),
[d]Max-Planck-Institut für Astronomie, Königstuhl 17, D-69117 Heidelberg, (Germany),
[e]LSW Heidelberg (Germany),
[f]Instituto de Astrofísica de Canarias (IAC), E-38200 La Laguna, Tenerife,(Spain),
[g]Instituto de Astrofísica de Andalucía, CSIC, Glorieta de la Astronomía s/n, 18008 Granada, (Spain),
[h]Centro de Astrobiología (CAB), CSIC-INTA), Carretera de Ajalvir km 4, E-28850 Torrejón de Ardoz, Madrid (Spain),
[i]Thüringer Landessternwarte Tautenburg (TLS) (Germany),
[j]Department of Physics and Astronomy, University of Florence, (Italy),
[k]INAF - Osservatorio Astrofisico di Arcetri, Largo E. Fermi 5, I-50125 Firenze, (Italy),
[l]INAF - Osservatorio Astronomico di Brera, Via E. Bianchi 46, 23807 Merate (LC), (Italy),

## ABSTRACT

We present here the preliminary design of the RIZ module, one of the visible spectrographs of the ANDES instrument[1]. It is a fiber-fed high-resolution, high-stability spectrograph. Its design follows the guidelines of successful predecessors such as HARPS and ESPRESSO. In this paper we present the status of the spectrograph at the preliminary design stage. The spectrograph will be a warm, vacuum-operated, thermally controlled and fiber-fed echelle spectrograph. Following the phase A design, the huge etendue of the telescope will be reformed in the instrument with a long slit made of smaller fibers. We discuss the system design of the spectrographs system.

**Keywords:** Ground-based instruments, High resolution spectrographs, Extremely large telescopes, Exoplanets, Stars and planets formation, Physics and evolution of stars, Physics and evolution of galaxies, Cosmology, Fundamental Physics, High Spectral Resolution.

# 1. INTRODUCTION

ANDES[1] is the high-resolution stable spectrograph for the ELT covering a wide spectrum from the U-band to the K-band. It is now in Phase B and PDR should be completed by mid-2025. ANDES is a suite of high resolution spectrographs that aim to provide high precision and high spectral accuracy. The design of ANDES inherits the design of several smaller instruments with similar requirements such as HARPS, ESPRESSO, NIRPS etc. ANDES is a single object (+ sky or calibration) fiber-fed instrument working either in seeing limited mode or in SCAO mode with an IFU (only for the YJH spectrograph in the baseline).

The RIZ spectrograph is one of the major subsystems ANDES[2]. Figure 1 shows the different parts of the instrument and their connections. The light is collected at the front-end[3] and transmitted to the different spectrographs via the fiber-link[4]. A calibration unit[5] provides the calibration light to the spectrograph through the fiber-link. The RIZ spectrograph subsystem start at the fiber slit that is the last part of the fiber-link and finishes at the detector.

In the baseline design the spectrograph will take the light from the seeing limited arm. As a goal we foresee the option to have an IFU[6] slit, in addition to the seeing limited slit, that could be fed by the SCAO[7] front end of ANDES or an other more performant AO system on the ELT, such as the future PCS[8].

The RIZ spectrograph will be located in the Coudé laboratory below the ELT as the fiber transmission losses allow for the ~ 150 m between the front-end and the spectrograph.

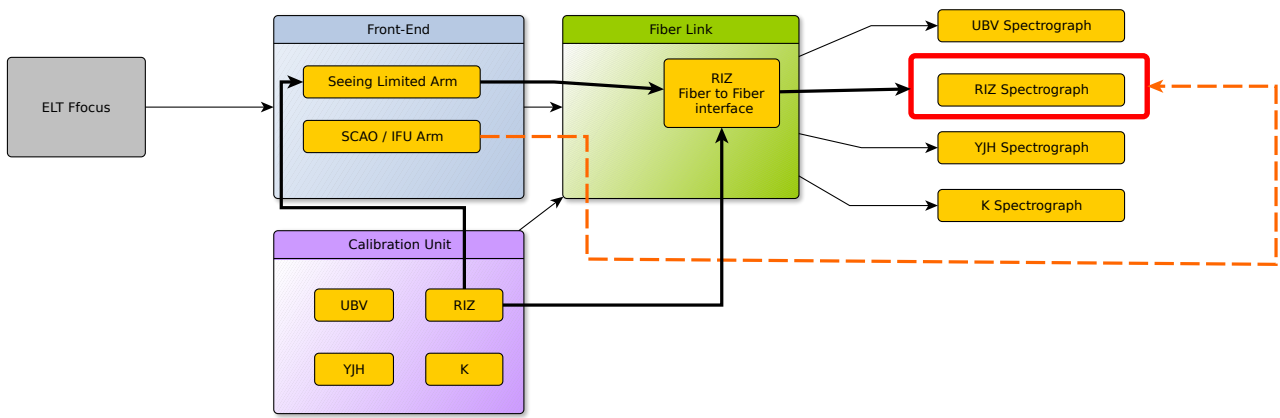

Figure 1: RIZ subsystem in the ANDES system. The RIZ spectrograph is connected to the rest of the instrument via the fiber link. The baseline design is to have only a seeing limited input to the spectrograph, but the design allows for the possibility to have in the future and additional input slit with an IFU mode (either using ANDES itself or another ELT AO system such as the future PCS).

Figure 2 shows the RIZ spectrograph specifically. Its design is strongly inspired from HARPS and ESPRESSO. It is fiber-fed, under vacuum and thermally controlled. The main products are : the opto-mechanical assembly, the vacuum control system, the detector control system, the vacuum control system, the thermal control system and the control software.

The other visible spectrograph of ANDES, the UBV module, shares a lot of commonality with RIZ thus to optimize the available resources, we share as many designs as possible. For example we have the same design for the vacuum tank and thermal enclosure, we share common design concepts for the optical and opto-mechanical designs, we have the same design for the detector heads and share the control system design.

The price to pay is that the common designs are based on the most constraining requirements as the spectrographs are in very different locations and environments.

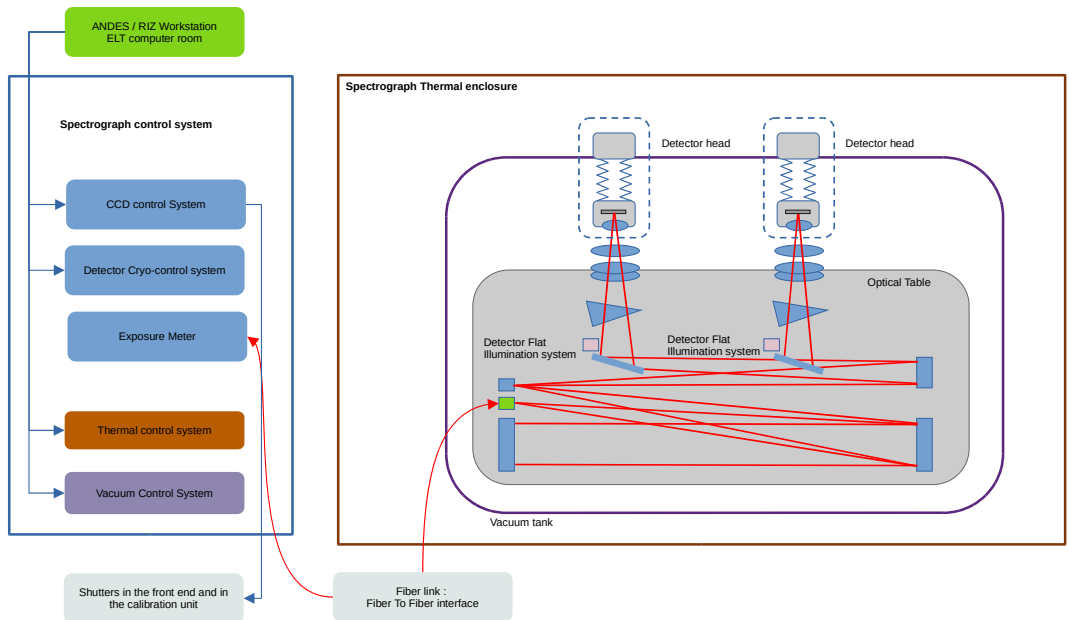

Figure 2: System scheme of the RIZ spectrograph showing most of the functional components.

## 2. MAIN DESIGN PARAMETERS

The spectrograph has to be stable at the level of 1 m/s goal 10 cm/s.

- The spectrograph has to be under vacuum with a pressure maintained bellow $10^{-4}$ mBar,
- The temperature stability should be better than 5 mK with a goal at 1 mK.
- The spectrum covered by the spectrograph is the R and IZ band.
- The spectrograph is located in the Coudé room of the ELT
- The fibers length to the spectrograph is ~ 150 m
- The input of the spectrograph is a slit with a total of 66 fibers, 31 for the object, 31 for the sky, two for simultaneous calibration and the rest are dark ones to allow for a proper separation of the fiber bundles.

## 3. DESIGN

The optical design[9] is very similar to ESPRESSO with some notable differences. The main one is that to cope with the étendue of the telescope and keep the instrument at a reasonable size the fiber coming from the telescope is sliced and reformatted in a very long slit. The resulting instrument is 1.5 times larger than ESPRESSO in all directions. It takes a large part of the Coudé room (see figure 3).

The spectrograph, like ESPRESSO, has 2 detectors (9Kx9K, 90 mm side) and thus 2 channels. One cover the R band from (620 to 760 nm) and the other the IZ band from (750 nm to 950 nm). The orders are elongated as they are composed of 66 fibers. There are 18 orders in the R and IZ cameras.

To meet the weight constraints of the Nasmyth platfom the common UBV/RIZ opto-mechanical design is based on aluminium, and in particular the optical table should be a welded structure out of the 6061-T6 alloy. The parts of the bench will be welded using Electron Beam Welding technology. Heat treatment at different stages of the construction will give the bench the stability required. Optical mounts are also made of aluminium and the interface with the optics, similarly to the ESPRESSO opto-mechanical design is based on glueing mechanical parts to the optics and having kinematic interfaces based on commercially available components such as Canoe Spheres* or Spherolinders† (clamping could be an alternative if some difficulty arises).

The vacuum chamber design[10], guided by the same constraints is based on cylindrical vessel made of aluminium. The thermal enclosure[10] design is more complex than for precedent similar instrument as we aim at a common design for Coudé and Nasmyth environment. We have a one layer design, using a large lightweight PUR enclosure with aluminium skins and control hardware.

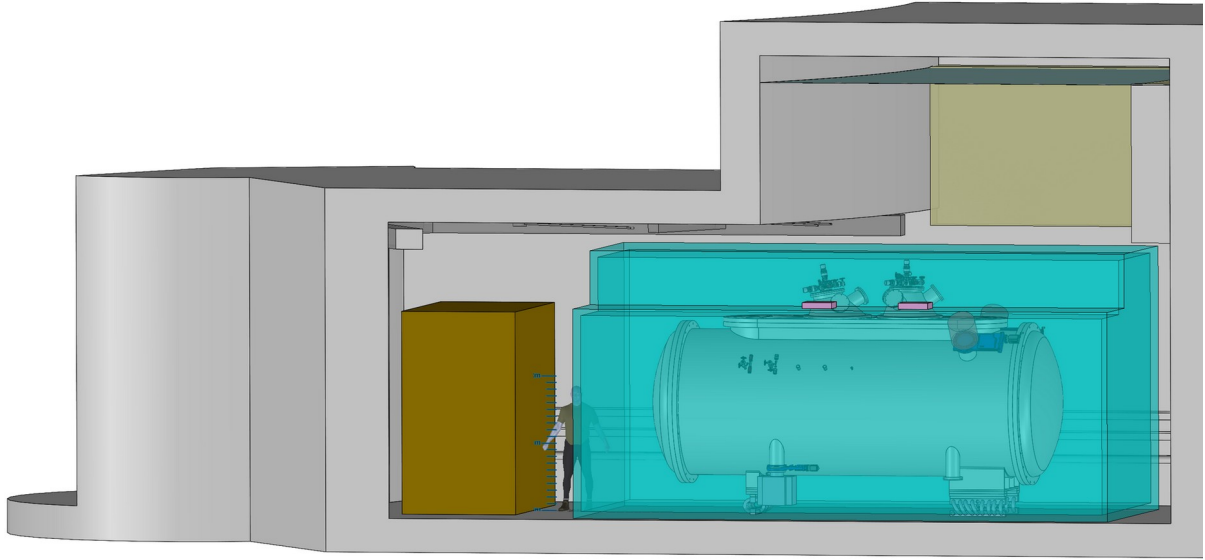

Figure 3: RIZ spectrograph installed in the coude lab at ELT.

The detector units[11] are a new development for ANDES, They are based on the classical differential vacuum cryostat of HARPS and ESPRESSO, where the detector is rigidly fixed to the optical table and with a soft bellow to the vacuum tank. The main change is the cooling system, passing from a Continuous Flow Cryostat, to a cryocooler version. This change is based on the experience gained for the N-ECAM camera on the EULER telescope and the detector head for RISTRETTO. This design is based on the ESO standard cryocooler Sun Cryotel GT. Prototyping activities are foreseen in the development to validate the concept for the larger detector that are used for ANDES.

The spectrograph control system consists of 3 subsystems (in order to map efficiently the distributed nature of the instrument development):

- Vacuum control system
- Thermal control system
- Detector control system

All are based as much as possible on the ESO standard hardware, and based on Beckhoff PLC systems.

The control software, essentially the detector control software and the device manager will be designed to adhere to the ESO ELT framework.

---

*https://www.precisionballs.com/All_Vee_Blocks.php#mid_size_vee
†http://www.g2-engineering.com/products-SPH

# 4. INTERFACES

The main external interfaces of the RIZ subsystems are the Coudé room and the SCP for network.

The main power, cooling and network connection will come from the internal ANDES distributors. All other interfaces are linked to the main function of the spectrograph.

The main optical interfaces are:

- <u>The front-end</u>: Dichroics between the spectral channels defining the RIZ bandwidth.

- <u>The fiber link</u>: The slit of the spectrograph is provided by the fiber-link subsystem. This interface is complex due to the high level of alignment precision required for the slit.

  The fiber-link provides the fiber-slit assembly that must pass through the thermal enclosure and the vacuum vessel and be mounted on the optical table of the spectrograph. The mechanical package of the fiber-slit as shown on figure 4 provides a mechanical protection to the fibers to allow for safe handling during integration. Several constraints comes with this slit:
  - We will align the system without the vacuum tank. We need to be able to remove the fiber-slit and reposition with little to no alignment.
  - We should be able in the future to be able to replace the single slit that is the baseline, to a potential IFU slit (a 2 slit unit, one for the seeing limited and one for the IFU).
  - We should be able to use a dummy slit for the integration period. It will provide at the same time a more independent workflow between subsystems, and also allow the system to be aligned and tested for the IFU case, i.e. having a second slit.

  Since the fiber slit has no mechanical reference to its optical axis, the solution is to have a subassembly that will provide an aligned beam to some mechanical reference. It consists of the fiber-slit and a pair of anamorphic lenses that deliver the beam to the main collimator at F/20 on one axis and F/10 on the other. It is called the Anamorphic Slit Assembly (ASA) and is shown on figure 4. Several ASA units will be produced adapted to the different slits.

  The different ASA units will be aligned to an alignment set-up so that the different assemblies can be interchanged on the optical bench.

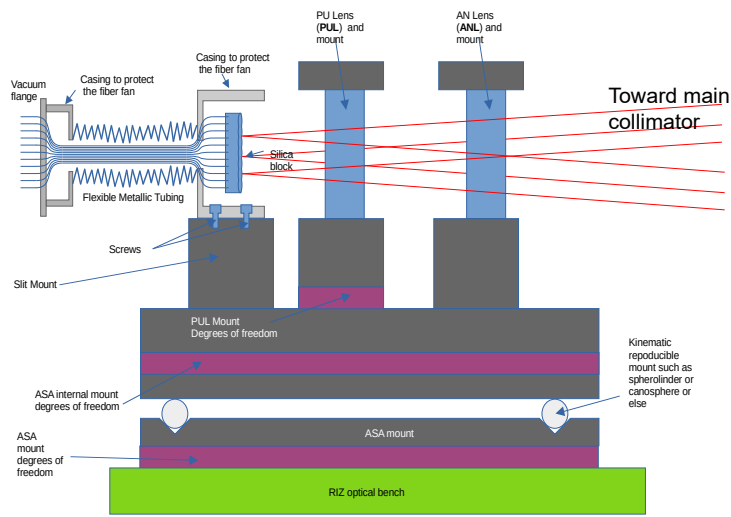

Figure 4: Scheme of the ASA unit.

The other important group of interfaces for the spectrograph are the shutters:

Being an ELT instrument the detectors are controlled and read using the ESO NGCII controllers. They are, among other things, in charge of the control of the shutters and of setting the time related keywords in the FITS header of the produced images. There are several shutters to be driven by the spectrograph: One in each seeing limited arm, one for the simultaneous calibration that does not pass through the front end, and finally two shutters for calibration beams when they pass through the front-end. Figure 5, shows how the shutters should be coordinated in the case of an object/sky with simultaneous calibration mode. The exposure meter should also be coordinated with the shutters.

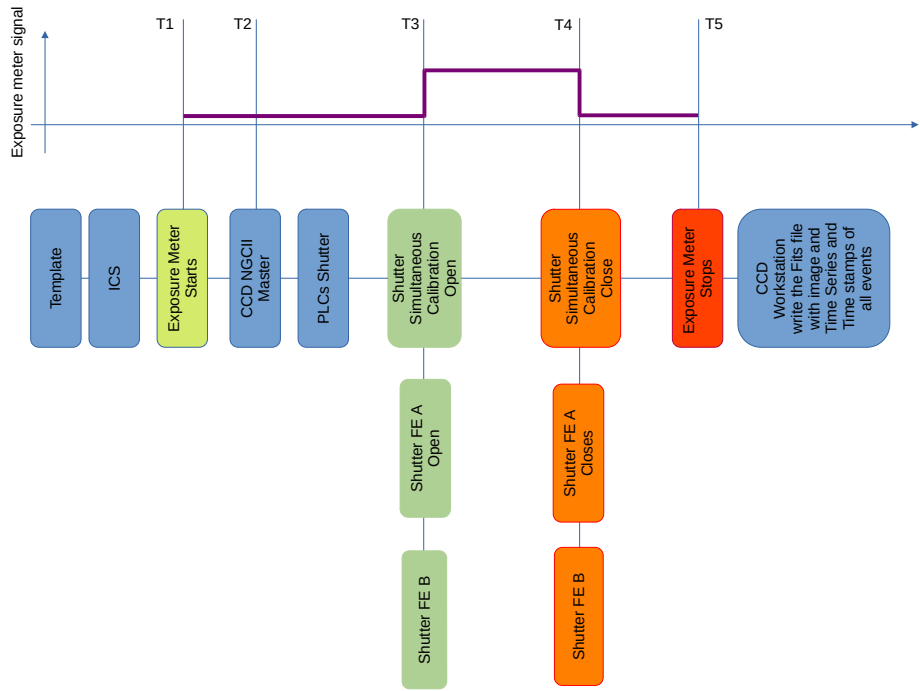

Figure 5: Scheme of the shutter / exposure meter coordination during and exposure with the configuration object/sky/simultaneous calibration.

The design of this system is not yet finalised, but possibly as in the case of ESPRESSO it will involve linking the NGCII controller to a PLC which will actually configure and control the various shutters operations.

## 5. PHOTOMETRIC PERFORMANCES

The spectrograph optics are numerous and we aim for an overall efficiency of the spectrograph, including the quantum efficiency of the detector, of 30%. The detectors are Teledyne CCD-290-99, a 90 mm detector, with 9Kx9K pixels. We are using the deeply depleted version, with the multi-astro-2 coating. This detector has over 80% of quantum efficiency in the R-band and drops from 850nm to 33% at 950 nm. This not optimal for the IZ-band, but is the result of a compromise in the distribution of the wavelengths across the instrument and the limits of the different detector technologies. To improve this, one would need a detector with a thicker silicon layer, deeply depleted versions are already 40 micron thick, some smaller detector are produced with layer thickness of 100 microns which has a large impact on the quantum efficiency of the detector. However this is at the expense of yield, cosmetics and performances. Such detectors are not available off the shelf for the size we need. We could consider making a mosaic with smaller detectors,but this has very important drawbacks. With the available thicker detectors one would require either 2x2 mosaic or 4x2. This would mean holes larget that 0.11 nm in all diffraction orders and partial loss of some orders.

On figure 6, we show the overall performance of ANDES RIZ spectrograph (when integrated in the ANDES system as a whole). The two requirements we have to fulfil are a transmission target (average better than 7%, and nothing lower than 4%). In the Z-band after 875 nm the efficiency of the system falls below the requirement, but without falling to 0.

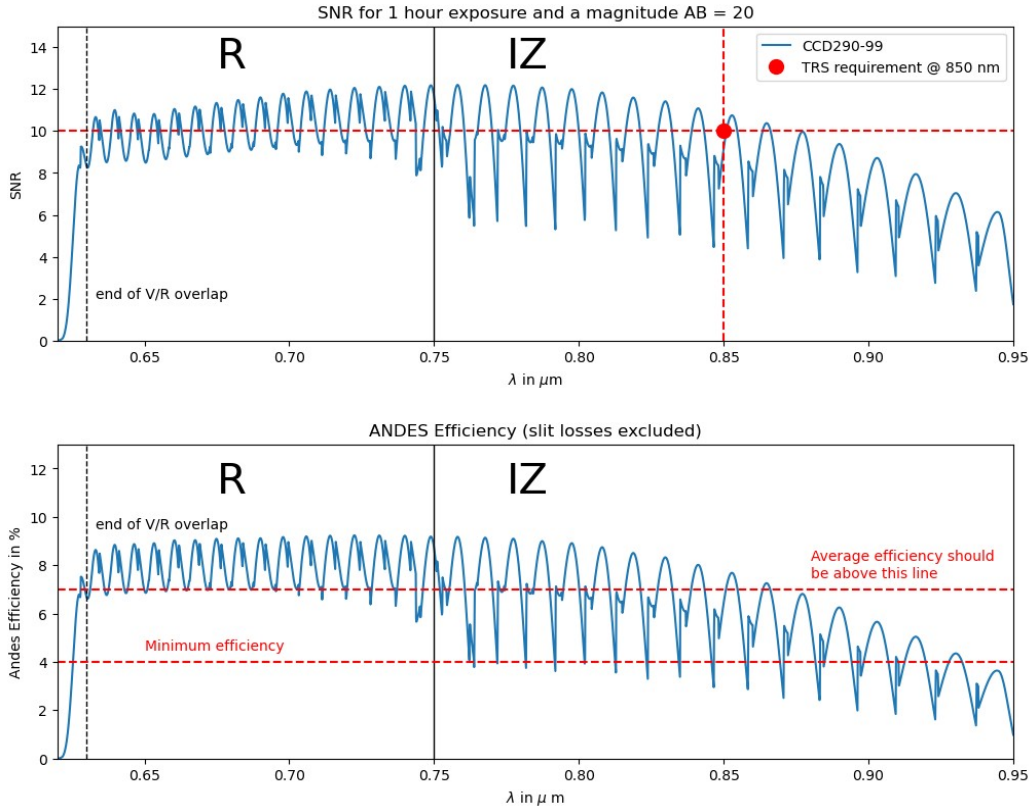

Figure 6: Photometric performance of the spectrograph. On top the SNR reachable in one hour of exposure for a source of AB Magnitude 20 and on the bottom the efficiency curve of the RIZ spectrograph inside of ANDES (slit loss excluded).

## 6. EXPOSURE METER

In order to provide an essential input for the barycentric correction to the data reduction software, it is necessary to measure the flux variation of the observed target to determine the mean time of exposure. There will be at least one exposure meter per spectrograph. We are still studying how many spectral channel per spectrograph are necessary. The highest accuracy necessary is 0.6 s for the mean exposure time, that would give a maximum error of 1 cm.s$^{-1}$.

To make the exposure-meter, the light is collected at the fiber-to-fiber interface, where the slicing of the beam from the large fiber coming from the front-end is taking place. Here a few percent of the light is lost anyway, leaving a total of six fibers for the object and six fibers for the sky to the exposure meter. The design is not ready yet but probably the exposure meter will be based on a technical camera.

## 7. INTERNAL FLAT ILLUMINATION SYSTEM

Another subsystem that needs to be provided is a direct illumination system for the detector from inside the vacuum chamber, in order to verify the calibration of the instrument over time. This has not been designed yet but, the implementation will probably be based on a LED installed in a strategic position for each arm.

## 8. CONCLUSION

We have presented in this paper the current state of the design of the RIZ spectrograph one of the major subsystem of ANDES. The preliminary design of the RIZ spectrograph will be completed at the system PDR, mid-2025[12]

## 9. ACKNOWLEDGMENTS

We would like to thank in particular Bernard Delabre that helped us with the spectrograph design concept.

We would like to thank the ANDES consortium and ESO to help us in this ambitious project.

The ANDES project is partially funded through the SNSF FLARE program for large infrastructures under grants

20FL21_173604, 20FL20_186177 and 20FL20_216577.

Manuel Amate Plasencia, Jose Peñate Castro, Afrodisio Vega Moreno, Fabio Tenegi, Roberto Simoes and Jonay I Gonzáles Hernández acknowledge financial support from the Spanish Ministry of Science and Innovation (MICINN) project PID2020-117493GB-I00.

Tobias M. Schmidt acknowledges the support from the SNF synergia grant CRSII5-193689 (BLUVES)

## 10. REFERENCES

[1] Marconi and al., "ANDES, the high resolution spectrograph for the ELT: science goals, project overview and future developments," Ground-based and Airborne Instrumentation for Astronomy X **13096–39**.
[2] Zanutta and al., "ANDES, the high resolution spectrograph for the ELT: model-based systems engineering approach," Modeling, Systems Engineering, and Project Management for Astronomy XI **13099–71**.
[3] Cabral and al., "ANDES, the high resolution spectrograph for the ELT: the front-end and its seeing limited arms," Ground-based and Airborne Instrumentation for Astronomy X **13096–164**.
[4] Tozzi and al., "ANDES, the high resolution spectrograph for the ELT: evolution of the fiber-link subsystem after the system architecture review," Ground-based and Airborne Instrumentation for Astronomy X **13096–170**.
[5] Huke and al., "ANDES, the high resolution spectrograph for the ELT: Calibration Unit(s)," Advances in Optical and Mechanical Technologies for Telescopes and Instrumentation VI **13100–27**.
[6] Tozzi and al., "ANDES, the high resolution spectrograph for the ELT: the integral field unit module," Ground-based and Airborne Instrumentation for Astronomy X **13096–173**.
[7] Pinna and al., "ANDES, the high resolution spectrograph for the ELT: design of the adaptive optics system," Adaptive Optics Systems IX **13097–129**.
[8] Kasper, M., Cerpa Urra, N., Pathak, P., Bonse, M., Nousiainen, J., Engler, B., Heritier, C. T., Kammerer, J., Leveratto, S., Rajani, C., Bristow, P., Le Louarn, M., Madec, P.-Y., Ströbele, S., Verinaud, C., Glauser, A., Quanz, S. P., Helin, T., Keller, C., et al., "PCS — A Roadmap for Exoearth Imaging with the ELT," The Messenger **182**, 38–43 (2021).
[9] Lanotte and al., "ANDES, the high resolution spectrograph for the ELT: RIZ spectrograph preliminary optical design," Ground-based and Airborne Instrumentation for Astronomy X **13096–167**.
[10] Santos Diaz and al., "ANDES, the high resolution spectrograph for the ELT: RIZ & UBV spectrographs' preliminary design, analysis, & integration of the Vacuum Vessel and Thermal Enclosure," Ground-based and Airborne Instrumentation for Astronomy X **13096–172**.
[11] Genolet and al., "ANDES, the high-resolution spectrograph for the ELT: RIZ and UBV spectrographs' preliminary design of the detector unit," X-Ray, Optical, and Infrared Detectors for Astronomy XI **13103–107**.
[12] Di Marcantonio and al., "ANDES, the high-resolution spectrograph for the ELT: project management for the preliminary design phase," Modeling, Systems Engineering, and Project Management for Astronomy XI **13099–74**.